# Preserving Location Privacy in Mobile Edge Computing

Hongli Zhang[1]    Yuhang Wang[1]    Xiaojiang Du[2]   and   Mohsen Guizani[3]

1 Research Center of Computer Network and Information Security Technology,
Harbin Institute of Technology, Harbin 150001, China,
Email：zhanghongli@hit.edu.cn, apple125110@gmail.com
2 Dept. of Computer and Information Sciences, Temple University, Philadelphia PA 19122, USA,
Email: dxj@ieee.org
3 Dept. of Electrical and Computer Engineering, University of Idaho, Moscow, Idaho, USA.
Email: mguizani@ieee.org

**Abstract.** The burgeoning technology of Mobile Edge Computing is attracting the traditional LBS and LS to deploy due to its nature characters such as low latency and location awareness. Although this transplant will avoid the location privacy threat from the central cloud provider, there still exists the privacy concerns in the LS of MEC scenario. Location privacy threat arises during the procedure of the fingerprint localization, and the previous studies on location privacy are ineffective because of the different threat model and information semantic. To address the location privacy in MEC environment, we designed LoPEC, a novel and effective scheme for protecting location privacy for the MEC devices. By the proper model of the RAN access points, we proposed the noise-addition method for the fingerprint data, and successfully induce the attacker from recognizing the real location. Our evaluation proves that LoPEC effectively prevents the attacker from obtaining the user's location precisely in both single-point and trajectory scenarios.

## 1. Introduction

Mobile Edge Computing (MEC) is a new technology which is currently being standardized in an ETSI Industry Specification Group (ISG) of the same name. Mobile Edge Computing provides an IT service environment and cloud-computing capabilities at the edge of the mobile network, within the Radio Access Network (RAN) and near mobile subscribers. The aim is to reduce latency, ensure highly efficient network operation and service delivery, and offer an improved user experience.The growth of mobile traffic and pressure on costs are driving a need to implement several changes in order to maintain quality of experience, the Internet of Things (IoT) is further congesting the network and network operators need to do local analysis to ease security and backhaul impacts [1].

Among all of the pervasive mobile and cloud-based services, the location-based service (LBS) and the localization service (LS) are the most suitable services for the decentralized deployment of the MEC scenario. On on hand, MEC is characterized by low latency, proximity and location awareness [2], these features of MEC are naturally fit with LBS and LS. On the other hand, in the vast majority of instances, the LBS and LS can be provided inside each subarea of MEC community independently, since most of the location-based information (e.g. the kNN queries, POI in proximity) are contained in the MEC scenario, data could be collected and processed based on location without being transported to cloud, this will block the LBS and LS provider from getting the location of user, which will preserve the location privacy better.

However, from the view of privacy preservation, in MEC scenario, the concern on the location privacy still exists. Although the location is avoided from being sent to the centralized cloud, the different threat model in MEC scenario is, to some extent, threatening the location privacy. As shown in Fig.1, two main factors are involved in the MEC scenario.

First, in the traditional threat model of centralized LBS service, the privacy concerns are originated from the geolocation information, and the preservation efforts are focusing on preventing the LBS providers from knowing the user's accurate locations while at the same time retaining the LBS functionality and service quality. However, in MEC scenario, such threat model is inapplicable, since the location awareness MEC is capable of generating location from the wireless signal space "fingerprint". In MEC scenario, this fingerprint is equivalent to the location from the view of privacy-preserving, and the fingerprint of user needs to be protected.

Second, the basis of the MEC infrastructure is constructed by the edge smart devices with limited computational power and with well-known lacking of security. As a result, although the centralized LBS providers which are considered to be untrusted (or curious-but-honest) are excluded, the privacy concerns due to the weak security of infrastructure is still severe.

In conclusion, we argue that the location privacy preservation in the MEC scenario should be performed in the very beginning of the generation of locations. The fingerprint information needs to be protected since it is equivalent to the location in MEC scenario. However, most of the state-of-the-art research on location privacy protection focuses on solving the privacy threat against LBS providers by investigating how to use the location safely or in a privacy-preserving way, they are incapable for the MEC scenario since the semantic content of wireless signal space fingerprint is different from the geolocation coordinates.

In this paper, we investigated the location privacy preservation in MEC scenario, and proposed a noise addition-based scheme named LoPEC to protect the fingerprint information of user. Specifically, we introduce the fundamental topology model of the

MEC wireless infrastructures, based on this model, we designed a method to generate the "noise fingerprint". Then, the noise addition scheme was given to protect fingerprint of the user. We consider the trajectory privacy and propose an enhanced algorithm, which can further generate trajectory-like noise fingerprints when using continuous location updates. To realize our scheme on smart devices directly without any additional system architectures, we propose a mechanism for the daily collection of a noise fingerprint candidate set. This mechanism also greatly enlarges the selectable range of noise fingerprint and enables the user to generate noise fingerprint beyond his current sensing range in real time.

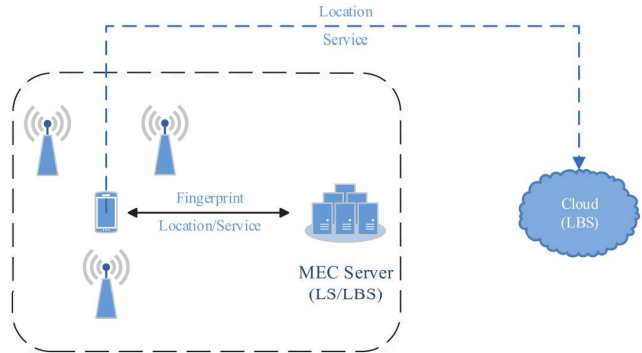

Fig.1 LBS and LS in MEC scenario.

Our approach has no negative impact on the functionality of the upper LBSs. The evaluation we implemented on Android device verified the effectiveness while maintaining a reasonable time cost for today's devices.

This paper makes the following main contributions.

(1) We propose a method for adding noise that can confuse the potential attackers and prevent it from recognizing the user's location in MEC scenario. The protection level can be adjusted according to the degree of the threat. With our method, the noise fingerprint does not reduce the usability of the real location and has no impact on the LBS functionality.

(2) We use a light-weight and realistic system architecture for the MEC environment. No unrealistic assumption or multi-party cooperation is needed. Our scheme can be realized directly in modern smart devices and the mobile internet ecosystem.

(3) We consider both single-spot positioning privacy and trajectory privacy and ensures, through the use of the noise fingerprint, that the attacker will generate trajectory-like locations that are not easy to decipher by committing a homogeneity attack [3].

(4) LoPEC provides an optimization algorithm based on the raw noise-generating method and drastically reduces the computational costs. This feature is meaningful because energy use is an important concern with mobile devices.

The rest of this paper is organized as follows. Section II briefly reviews the related work. Section III outlines the preliminaries of LoPEC. We describe the design details of LoPEC in Section IV. Section V and Section VI provide theoretical analyses and experimental evaluations. Finally, Section VII concludes the paper.

## 2. Related work

Approaches on location privacy can be divided into two categories: LBS scenarios and LS scenarios.

## 2.1 Approaches to the LBS scenario (how to use safely)

Much attention has been paid to this location privacy scenario in the past decade. Most approaches to this scenario were surveyed comprehensively in [3][4][5][6][7].

The threat models of these approaches are more or less similar. They prefer treating the LBS providers as the most untrusted adversaries with extensive background knowledge. Based on this framework, their primary purpose is to prevent the LBS providers from knowing the user's accurate locations while at the same time retaining the LBS functionality and service quality as much as possible.

Many ingenious methods have been adopted successfully in these approaches, such as dummy adding, k-anonymity, obfuscation, region cloaking, caching and encryption-based methods. Although they have different ideologies and realization details, these approaches share the same understanding of the scenario boundary; they do not care how a location is obtained. The mobile user first obtains a known and definite location, and then their approaches are applied.

In MEC scenario, the LS providers (LP), who are as untrustworthy as the LBS providers, suffer from the same privacy threat concerns. The location privacy needs to be protected even before the location is generated. However, these approaches can neither protect location privacy against the LP because their threat model is incomplete, nor be applied to the MEC scenario directly because the communication contents as well as the semantics are different.

## 2.2 Approaches to the MEC scenario (how to obtain safely)

Compared to the LBS scenario, the situation in which the device requests the location from a LP is quite different. Several researchers have studied the problem in this scenario from different perspectives.

Damiani and Cuijpers described this privacy issue in [9] and tried to solve this problem by designing a policy control mechanism to adjust the granularity of the location determined by the LP. This approach is not a computational technique but rather a policy suggestion.

[10] Described a location privacy threat called a location spoofing attack, in which the attackers can counterfeit the device's original fingerprint information and imitate the real user to request the user's location from the LP. Then, the method proposes a reliability determination algorithm to cope with this threat. However, their threat model regarded the

LP as trusted, which is unrealistic in the MEC scenario.

Some encryption-based method such as [11] are also proposed to protect location privacy against the LP. However, the encryption procedure is too time-consuming for the lightweight edge devices and key management [12][13][14] is also a challenge.

## 3. PRELIMINARIES

In this section, we first introduce the assumptions on the system adopted in this paper using some basic concepts and then present the motivation for and the basic idea of our solution.

### 3.1 System assumptions

Our threat model is concise. We concentrate our research on the LS procedure in the MEC scenario, and we do not consider the location privacy problems of the LBS scenario. The LS providers are the most direct and shrewdest adversaries in our work. They may record people's locations or even daily trajectories without permission. Moreover, we consider the communication channel between device and LP to be trusted since this channel can be well protected by security protocols. Furthermore, attacks by exploiting the vulnerabilities of the device's operating system and programming framework [15][16] are beyond the scope of this paper.

Second, multiple dominant positioning technologies are being utilized by today's mainstream edge computing-based localizations, but we do not assume on what specific technology may be adopted by LS providers because our scheme is designed based on the fundamental principles of the vast majority of those technologies.

Finally, we treat the RAN access points (APs) as the only communication infrastructure adopted by the LP. This assumption is rational because state-of-the-art technologies mainly rely on the RAN APs (e.g. Wifi hotspots) to achieve meter-level positioning accuracy; other infrastructures such as cellular towers are only subsidiaries. However, our approach can be simply transplanted to other infrastructures such as cell towers because their basic positioning fundamentals are almost the same.

### 3.2 Limitations

Our goal is for our scheme to possess strong suitability for today's positioning technologies rather than apply to one specific positioning technology; we also want it to be realistic. This goal imposes the following limitations:
    1. LoPEC cannot rely on the technical details of any positioning technology; and
    2. LoPEC cannot perform any modification of existing positioning technologies.

To accommodate these limitations, we implement LoPEC on the device side. This approach can also simplify the design complexity of LoPEC because no matter what positioning technology the LP adopts, the user's job during the positioning process remains almost the same: sense the APs in proximity and transmit them to the LP (Fig.2(a)).

As we noted in Section 1, our basic idea is to use a noise-addition method to prevent the LP from knowing the user's accurate location. This idea is easy to implement in the LBS scenario because the location coordinate's semantic is uncomplicated and easy to simulate, even using the simplest randomization method, people can still ensure the equivalence between the real location and the noise one, all left people to do is to improve the noise fingerprint's degree of similarity in other aspects.

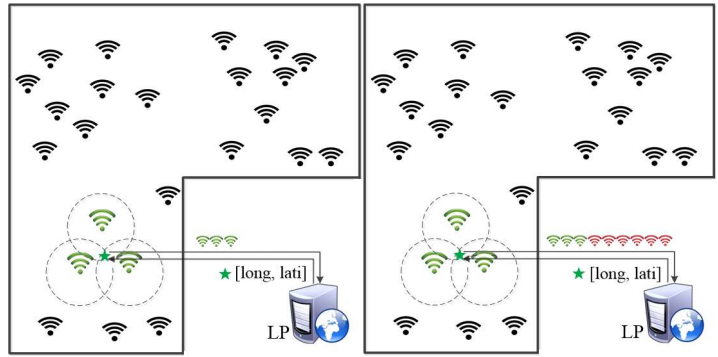

(a) initial way of LS without any protection     (b) using fabrication method to generate noise fingerprint

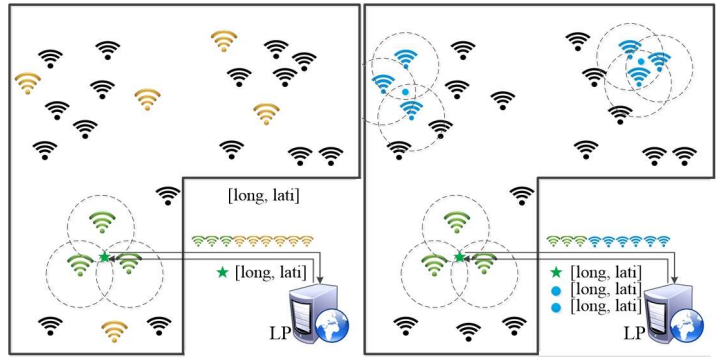

(c) using real world APs but with spatial distribution as noise fingerprint     (d) using cluster-like APs as noise fingerprint wrong (our basic idea)

Fig.2 Our basic idea of noise adding

However, the same concept of noise addition is much more difficult to realize in the MEC scenario for the following two reasons:

1. The noise fingerprint we need in the MEC scenario, i.e., the AP identifiers(Wi-Fi Mac addresses in this paper), must not only be homogeneous to real fingerprint but also should be understandable to the LP. As shown in Fig.2(b), if we use the fabrication method, which researchers used in the LBS scenario to manufacture dummies, the LPs can easily recognize the real fingerprint because the noise fingerprint does not exist in their databases.

2. Even using real-world AP identifiers that can be handled by the LPs as our noise fingerprint, there is still an additional problem to be solved; Fig.2(c) illustrates this problem. According to the characteristics of smartphone sensing, the spatial distribution of the real APs should satisfy a special cluster-like pattern; more precisely, the coverage areas of APs in the actual fingerprint must overlap one another. The features of the distributions of the real fingerprint and the noise fingerprint ought to be the same, which makes adding the noise more difficult.

## 3.3　Basic idea

For the first problem, we will enable the user to possess a huge number of real-world AP identifiers as the candidate set of noise fingerprint. For practicability, we will not assume a third-party broker, which possess the entire region's AP identifiers, would handle this problem. This assumption is convenient for simplifying the problem but not realistic for the implementation. In this paper, based on the fact that a device can sense and record a considerable number of APs during daily activity and movement, we design a Self-Collect and Self-Organize Algorithm (SCSOA) for smartphones to collect APs as the candidate set of noise fingerprint (Section 4.2).

For the second problem, we generate noise fingerprint with high similarity to the real fingerprint. Fig.2(d) shows this optimal case, when the noise fingerprint is difficult to distinguish. This process requires us to model the overlapping relations between APs in intuitive and predictable way and to make sure the smartphones can learn this model.

We use an undigraph to characterize the spatial distribution and the overlapping relations between APs. We first simplify the irregular cover area of the APs to a circular area (Fig.3(a); this simplification does not affect our model's authenticity because it is only for convenience of the display). Then, we can model the APs' spatial distribution as an undigraph G(V, E), in which Vertex(V) represents all APs in a region and Edge(E) represents the situation in which two APs in V have an overlapping relation in the spatial distribution.

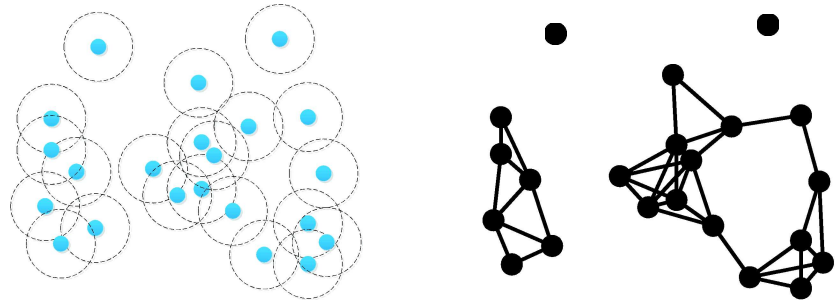

(a) Spatial distribution of APs with their "overlapping" relations　(b)Modeling result based on our rule

**Fig.3 Use undigraph to model APs spatial distribution**

Fig.3(b) illustrates the modeling result of the data shown in Fig.3(a). Note that G does not contain the geo-location coordinates of the APs; it only characterizes their spatial topology. Using this model, we see the overlapping spatial relations of the APs in real fingerprint can now be modeled as a complete subgraph in G. Then, with the help of G, we can generate noise fingerprint that has the same features as a real fingerprint; more specifically, we find some other complete subgraphs in G and use the corresponding APs as our noise fingerprint.

Because the SCSOA can generate G for device iteratively while sensing and recording APs, whenever a user requests LS from the LP, we can calculate several complete subgraphs from G without assistance from another party. However, restricted by the high real-time performance and low energy consumption demands of the device, the raw graph traversal algorithm for complete subgraph discovery is too time-consuming to apply, especially when G is large. To overcome this problem, we propose a faster Complete Subgraph Discovery Algorithm (CSDA). CSDA is based on the notion of clustering coefficients. The clustering coefficient of a vertex in a graph quantifies how close its neighbors are to being a clique (complete graph). CSDA tends to find APs associated with high clustering coefficients and is an efficient method, as demonstrated in Section 4.3.

Finally, we further consider the trajectory privacy in which the noise-addition method will suffer from the homogeneity attack. We protect against this threat by enabling the smartphones to generate noise fingerprint with continuous spatial distributions. Our basic idea is to find a complete subgraph adjacent to the previous one as best as we can (Section 4.4).

## 4. OUR PROPOSED SCHEMES

We first introduce an overview of our approach, and then we present our AP collection algorithm (SCSOA) and the noise fingerprint generation algorithm (CSDA) in detail. Finally, we present the enhanced CSDA (e-CSDA) for the trajectory scenario to defend against the homogeneity attack.

## 4.1 System overview

In our system, when a user requests LS in a fixed location, he first uses CSDA to generate several noise sets of APs as the noise fingerprint and then mixes them with the real sensed APs and sends them to the LP. CSDA runs on the undigraph G, which is created by SCSOA. SCSOA runs continuously in the background to collect APs and model G during the user's daily movements. It is evident that the longer this algorithm operates, the larger the undigraph G will be, which means a larger set of noise candidates and better privacy protection.

When the user is in the trajectory scenario (e.g., using a navigation LBS), the system will activate the e-CSDA to generate noise fingerprint with continuous spatial distribution to further strengthen the privacy protection.

## 4.2 SCSOA for APs collection

Inspired by the relationship between the APs' spatial distribution and their corresponding undigraph model, we can assert that given a set of APs with known spatial distribution and their undigraph G, if a user senses his surrounding APs, denoted as $AP_{real}$, then the vertices in G corresponding to AP$_{real}$ must constitute a complete subgraph.

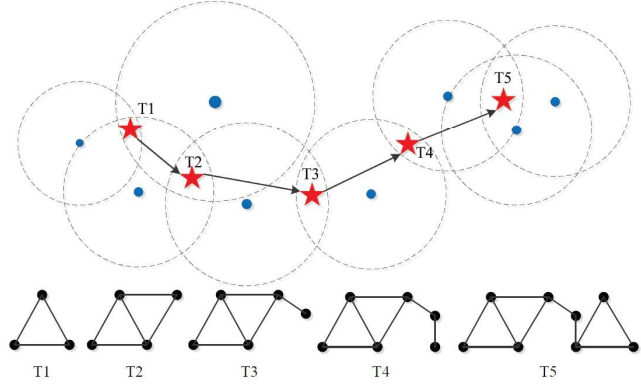

**Fig.4 Basic idea of SCSOA, user travels from T$_1$ to T$_5$ along the path shown as the black arrow and uses SCSOA to collect APs and generates G each time he senses some new APs**

Based on this feature, our main concept is to construct a complete subgraph and merge it with the existing undigraph each time the user senses and records his surrounding APs. By repeating this step, the user can construct a large enough undigraph for noise fingerprint generation; see Fig.4. Algorithm 1 details this process.

---

**Algorithm 1** Self-Collect and Self-Organize Algorithm (SCSOA)

**Require:**
    current undigraph $G(V,E)$;
    vertexes of current sensed APs $AP_{real}$;
    signal strength threshold $\tau$;

**Ensure:**
    $G$;

1: $V = V \cup AP_{real}$;
2: **for** each $v_i, v_j$ in $AP_{real}$ **do**
3:    **if** $e(v_i, v_j) \in E$ **then**
4:      return;
5:    **else**
6:      **if** $ss_i \geq \tau$ and $ss_j \geq \tau$ **then**
7:        $E = E \cup e(v_i, v_j)$;
8:      **end if**
9:    **end if**
10: **end for**

---

In real-world situations, an AP's signal strength may vary with time and environment and hence make its cover area unstable, especially in the boundary area. To improve G's authenticity, we use a threshold τ of signal strength to filter out those unstable overlapping relations. We first let V absorb the newly discovered APs, and then we add an edge into E selectively, according to τ: an edge will be in E if the signal strengths of both of its two vertices are greater than τ. Note that larger τ will lead to a higher authenticity of G but a smaller number of candidates for noise fingerprint.

SCSOA enables the user to construct an undigraph model of real-world APs as the candidate pool for noise generation. It also makes it possible for the user to implement LoPEC on his own, without any third-party participation. This capability further improves our system's practicality.

## 4.3 CSDA for noise-data generation

Our goal is to find a complete subgraph from G. This problem can be classified as a Clique Problem. However, the brute-force algorithm for finding a clique in an undigraph is too time-consuming for the LS scenario. Although this brute-force search can be improved by using more efficient algorithms, all of these algorithms require exponential time to solve the problem [17]. This requirement limits the utilization of this technique in our system in cases where G is large.

We solved this issue by dividing the whole problem into two phases. First, we determine the clustering coefficient c of each vertex in G and record the coefficients during the user's idle time (i.e., when the user is not in the LS scenario). As c can describe the closeness of a vertex's neighbors, the neighbors of a vertex with higher c are more likely to constitute a complete subgraph (with the edges between them); when $c_i$, we can be sure that $v_i$ and all its neighbors would be perfect candidates for use as noise fingerprint. Based on this feature, in the second phase (LS scenario), we adopt a randomization method to select noise fingerprint near the vertices with very high c (approximately 1). Algorithm 2 describes the details of this method.

---
**Algorithm 2** Complete Subgraph Discovery Algorithm (CSDA)
---
**Require:**
  current undigraph $G(V,E)$;
  vertexes of current sensed APs $AP_{real}$;
  clustering coefficient threshold $\epsilon$;
  number of noise fingerprint set $h$;
**Ensure:**
  $h$ noise fingerprint sets $AP_{noise}^1, \cdots, AP_{noise}^h$;
1: $V_h$= { vertexes with $c > \epsilon$} and $n$= number of vertex in $AP_{real}$;
2: randomly select $h$ vertex out of $V_h$ as candidate $V_c$;
3: **for** each $v_i$ in $V_c$ **do**
4:   **if** $n \leq$ (number of $v_i$'s neighbors) **then**
5:     randomly select $n$ vertexes out of $v_i$'s neighbors in $G$ as a noise fingerprint set $AP_{noise}^i$;
6:   **else**
7:     reselect $v_i$ from $V_h$ and back to step 4;
8:   **end if**
9: **end for**
---

We use those vertices with c higher than a threshold $\epsilon$ as the pointers to the dense areas of G. In this way, we avoid those time-consuming deterministic algorithms and perform a randomize method instead. We first randomly locate h dense areas in G and then randomly select several vertices as the noise fingerprint set in each area. Here, we further consider the fact that the number of APs in $AP_{real}$ may be different each time the user requests LS because of the unstable signal environment. Therefore, to reinforce the indistinguishability, the noise fingerprint must not only constitute a complete subgraph in G but must also contain the same number of vertexes as $AP_{real}$. Thus, we will deprecate those vertices with fewer neighbors than the real fingerprint.

Benefitting from the two-phase design and the randomization, CSDA can generate noise fingerprint far faster than deterministic algorithms. Furthermore, by randomly locating dense areas in G, CSDA can ensure the even distribution of the noise fingerprint selection from the probabilistic perspective, and randomly choosing vertices from the dense area will enrich the diversity of the noise candidates.

Note that CSDA will sometimes generate a false noise fingerprint set (i.e., not a complete subgraph in G) due to the situation when the coefficient c of the pointer vertex (in Vh) is not 1. However, this situation occurs with very small probability, and, as we show in our evaluation, this probability can be quantifiably controlled by adjusting ϵ. Because the SCSOA functions continuously, this probability will be further reduced.

## 4.4 e-CSDA for the trajectory scenario

We take into consideration the trajectory privacy in which the attacker (LP) can carry out a homogeneity attack to identify the user's real location. The homogeneity attack is mainly based on the fact that in a continuous LS request process, the locations should be homogenous with each other in spatial distribution. The attacker can distinguish the real user from the low-homogeneity noise fingerprint.

We enhance the CSDA and use noise fingerprint that possesses the trajectory-like spatial distribution to overcome this problem. Algorithm 3 illustrates this enhancement.

---
**Algorithm 3** enhanced CSDA (e-CSDA)
---
**Require:**
    vertexes of current sensed APs $AP_{real}$;
    noise fingerprint sets from last CSDA or e-CSDA $AP_{noise}^1, \cdots, AP_{noise}^h$;
**Ensure:**
    $h$ new noise fingerprint sets $AP_{noise}^1, \cdots, AP_{noise}^h$;
1: $c_j$ is the cluster coefficient of $v_j$;
2: $n_j$ is the number of $v_j$'s neighbors;
3: $n$ is the number of vertex in $AP_{real}$;
4: **for** each $AP_{noise}^i$ ($i$ from 1 to $h$) **do**
5:    randomly select a vertex $v_j$ from $AP_{noise}^i$;
6:    **if** $c_j \geq \epsilon$ and $n_j \geq n$ **then**
7:      randomly select $n$ vertexes out of $v_j$'s neighbors in $G$ as a noise fingerprint set $AP_{noise}^i$;
8:    **else**
9:      reselect $v_j$ from $AP_{noise}^i$ and back to step 6;
10:    **end if**
11: **end for**
---

We use the vertex adjacent to the last generated noise fingerprint as the start of our algorithm instead of using randomization every time. In this way, we can generate noise fingerprint neighboring the previous noise fingerprint. This process is rational because, although G does not contain the spatial information of the APs, two adjacent APs in G are very likely to be close to each other in the general case.

## 5. ANALYSIS

We provide our privacy metric to measure how much privacy our system can offer during an LS process. Then, we discuss several security issues.

## 5.1 Privacy metric

The number of noise fingerprint has been widely adopted as the privacy metric in noise addition-based approaches that are not restricted to location privacy protection. In our system, the number of noise fingerprint represents how many APs we use to mix with the real APs. These noise fingerprint are used to generate fake locations for the LP; therefore, we use the number of locations that can be determined using the noise fingerprints as our privacy metric, which can be defined as

$$H = \frac{k}{n}$$

Here, H denotes our privacy metric, k denotes the total number of noise APs, and n is the number of APs in the real fingerprint. According to our noise-generation algorithms, H will always be an integer, and higher H provides a higher privacy degree.

## 5.2 Security discussion

As we noted in Section 3.1, the LP can determine the user's location during the LS process. In our system, the noise-addition method generates high-similarity noises, which means that (1) these noise APs are all obtained from the real-world collection and (2) the LP can use these noises to calculate locations in the same way that it uses the real fingerprint. Thus, ideally, given the privacy metric h, the LP cannot distinguish the real location from the other h locations generated by our noise fingerprint sets, and the probability of a "luck guess" is $\frac{1}{h+1}$.

However, the LP may perform a distribution attack [3] based on his knowledge of the query probability of each AP in the whole area. The query probability of an AP indicates how frequently this AP was used as the search condition (real fingerprint), and this number is different for each AP because of the diversity of their spatial distributions (e.g., a public Wi-Fi-router in a mall will be used more frequently than a private home-edition router). The LP could use this information to narrow the guessing range. In our system, this disturbing knowledge will be gradually degraded because of our randomized algorithms for choosing APs. After a period of system implementation, the query probability of each AP will lead to equalization and the LP will lose this weapon against the user.

Finally, when in the trajectory scenario, the LP could carry out a homogeneity attack to infer the user's location from a continuous positioning request. In Algorithm 3, we have ensured that the location generated by our noise fingerprint has the same spatial continuity as the real fingerprint. It must be pointed out that we did not take the direction and velocity features of real location data into consideration because G cannot offer any spatial information for us to perform such an optimization. Thus, our noise trajectories may not have the same high-level features as the real trajectory. However, we still successfully strengthen the discrete noise locations and organize them in a trajectory-like spatial distribution. As a result, it is still difficult for the LP to recognize the real trajectory.

## 6. EVALUATIONS

LoPEC was evaluated by implementation on a smartphone with the Android 5.0 system, and the associated LP and LS were realized by simulation on our server. We simulate the MEC location service and the AP information used to calculate the location by the LP was collected from an urban area in the city Harbin China. We considered four aspects in our evaluations.

## 6.1 Simulation and implementation

As the fundamental data of our evaluation, we gathered the AP data (Wi-Fi Mac address and signal strength) from the real world by installing a simple program on a smartphone and using it to sense the surrounding Wi-Fi information. Fig.5 shows the area and the coordinates we gathered with Google Earth.

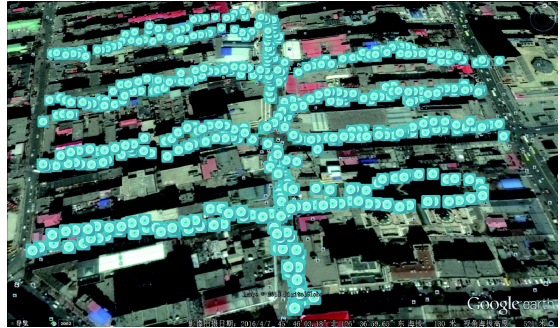

**Fig.5 Area and the coordinates we gathered APs from**

We chose a Wi-Fi-rich downtown business district to gather our data to obtain a better experimental result. For each coordinate in Fig.5, we recorded its geographic coordinate (in the latitude-longitude frame) and the surrounding Wi-Fi Mac addresses with their signal strengths (mean of 10 times for each), and we gathered 2016 Wi-Fi routers. For the convenience of data analysis and evaluation, we transformed the geographic coordinates into the Cartesian coordinate system using the Gauss-Kruger projection.

Based on the above data, we simulated the LP on our server to provide LS for user. To verify the universality of LoPEC, we realized two different positioning technologies, introduced in [18] (RADAR) and [19] (PBL), on the server side. These two classic approaches outlined the fundamentals of today's third-party RF-based positioning services.

We realized and implemented LoPEC on a Samsung smartphone with the Android 5.0 system . LoPEC runs in the background of the OS and protects the user's location privacy when an LS is used.

## 6.2 Quality of generated G

We invited volunteers to collect Wi-Fi information for us by using the SCSOA. SCSOA

ran in the background of their smartphones and generated G gradually as they shopped in the area shown in Fig.5. To evaluate the quality of G, we compared G with $G_t$ to observe the change in quality as the number of volunteer trips increased. Here, $G_t$ is the global undigraph of the area shown in Fig.5. We obtained this value using all the Wi-Fi information gathered in this area as the input of SCSOA. Obviously, a higher similarity between G and $G_t$ means that G is of higher quality. Fig.6 shows the influence of the travel time t on the quality of G.

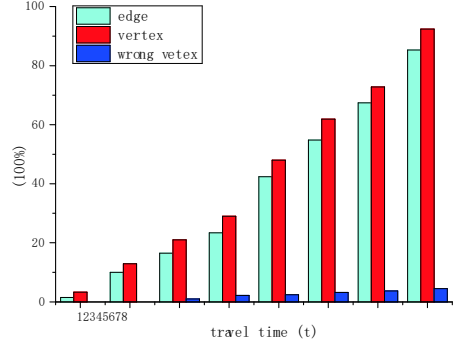

Fig. 6 Quality of generated G vs. travel times

We studied $\frac{|V(G)|}{|V(G_t)|}$, $\frac{|E(G)|}{|E(G_t)|}$ and the number of incorrect vertices (vertices that do not belong in G) as the travel time increased from 1 to 8. The result shows that SCSOA can generate G efficiently over time. Both the Wi-Fi-routers (vertices) and their overlapping relations (edges) can be well modeled in G. Moreover, the ratio of incorrect vertices in G is fairly small, and these incorrect vertices are probably due to the unfixed Wi-Fi-routers (e.g., the portable smartphone hotspots) and the Wi-Fi-routers beyond our measured boundary.

## 6.3 Success rate of clique noise

We evaluated the impact of the following four parameters on the success rate of our CSDA: privacy metric h, clustering coefficient threshold ϵ, the scale of G and different positioning technologies. Fig.7 shows the evaluation results.

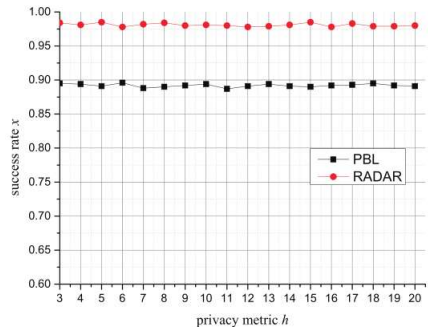

(a) x vs. h with ϵ = 0.95

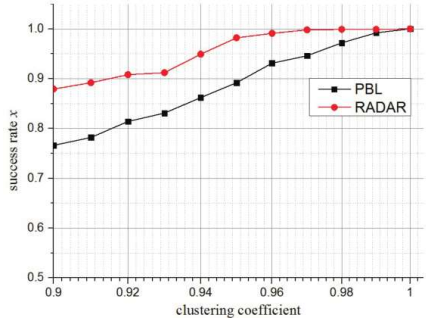

(b) x vs. ϵ with h = 1

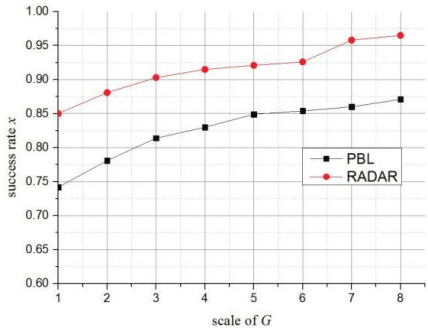

(c) x vs. scale of G with ϵ = 0.95, h = 1

Fig.7 Effects on success rate

The success rate x is independent of the privacy metric h (Fig.7(a)) because each noise fingerprint is generated independently from the others in CSDA. This feature is inspiring when the user needs to promote his privacy metric. x increases when the clustering coefficient threshold is tightened (Fig.7(b)), as higher ϵ means more strict and dense area filtration. In addition, x will increase as G enlarges due to the rise in G's density and clique number (Fig.7(c)). Furthermore, we find that x can remain at a high level (85% of RADAR in our experiment) even on the smallest G in our evaluation. This result confirmed that LoPEC could function quite well for various types of users.

The x on RADAR outperforms x on PBL. This result is caused by the differences between these two positioning technologies in their data format requirements. More specifically, RADAR tends to use a fixed number (usually less than 5) of APs to calculate the coordinate; in contrast, PBL will utilize as many APs as it can. This leads to the fact that CSDA always has to generate a larger complete subgraph for PBL than for RADAR, and this will influence x. In general, CSDA can obtain a considerably high success rate in various conditions.

## 6.4   Computational cost

As SCSOA runs in the background in the user's smartphone and does not affect the LS experience, we focus our attention on CSDA and e-CSDA due to their extra time delays.

Before our evaluation, we first introduce the brute-force algorithm for comparison. The brute-force algorithm will find an m-clique in an n-vertex graph by checking all $C_n^m$ m-subgraphs for completeness. Fig.8 shows the effects of h and different scales of G on the CPU time of each algorithm.

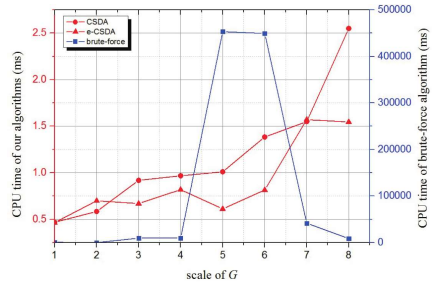

(a) t vs. scale of G with h = 5 on PBL

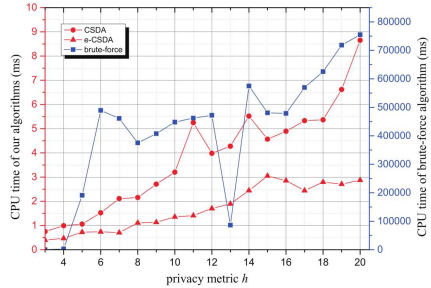

(b) t vs. h on PBL

Fig.8 Computational cost

In Fig.8(a), we set the privacy metric h to 5 and select 8 incremental subgraphs from the previous evaluation. The brute-force algorithm takes a considerably longer and very unstable amount of CPU time. In contrast, the computational costs of our CSDA and e-CSDA are lower by more than an order of magnitude than that of the brute-force algorithm and increase linearly with the size of G. The same result is shown in Fig.8(b), where we perform these algorithms on a certain G and increase the privacy metric h from 3 to 20. In addition, our algorithms perform better on RADAR than PBL for the same reason as described in the above subsection. The result shows that LoPEC utilizes the randomization method successfully, and its computational cost is acceptable for the smartphone LS scenario.

## 6.5 Trajectory feature of noise fingerprint

We evaluated the trajectory feature of the noise fingerprint by performing a positioning experiment. We walked through the area shown in Fig.9 in a casual path (green) and sent a series of RF data, which was protected by our e-CSDA, and we used RADAR to calculate the location of each dataset. Fig.9 shows the result of this evaluation, and we can see clearly that our noise fingerprint (blue) has a trajectory-like spatial distribution, just like the real fingerprint.

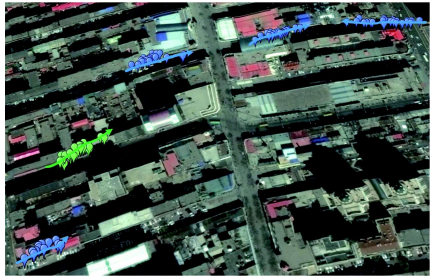

Fig.9 Spatial distribution of Trajectory-like noise fingerprint generated by e-CSDA

## 7. Conclusion

This paper proposes LoPEC, a location privacy-preserving scheme for the mobile edge computing scenario. We argue that the preservation of location privacy in MEC is equivalent to the protection of the wireless fingerprint. Based on the good modeling of the AP spatial distribution, we found a way to generate high-quality noise fingerprint. SCOSA provides the user with a self-sufficient way to apply LoPEC without the assistance of a broker. Then, we proposed two randomization-based noise-addition algorithms: CSDA and e-CSDA. CSDA greatly reduces the computational cost of the raw noise generation method by utilizing the notion of the clustering coefficient and still retains a high success rate. The e-CSDA further protects the location privacy in the trajectory scenario. Evaluation results indicate that LoPEC can protect the user's location privacy in MEC environment.

## 8. Acknowledgement

This work was supported by National Natural Science Foundation of China (Grant No. 61723022, 61601146), and the National Key research and Development Plan (Grant No. 2017YFB0803300).

## References

[1]     Hu Y C, Patel M, Sabella D, et al. Mobile edge computing—A key technology towards 5G[J]. ETSI white paper, 2015, 11(11): 1-16.
[2]     X. Du and H. H. Chen, "Security in Wireless Sensor Networks," IEEE Wireless Communications Magazine, Vol. 15, Issue 4, pp. 60-66, Aug. 2008.
[3]     WANG Y H, ZHANG H L, YU X Z. Research on location privacy in mobile internet[J]. Journal on Communications, 2015, 36(9): 2015167.
[4]     Chow C Y, Mokbel M F. Trajectory privacy in location-based services and data publication[J]. ACM Sigkdd Explorations Newsletter, 2011, 13(1): 19-29.
[5]     Krumm J. A survey of computational location privacy[J]. Personal and Ubiquitous

Computing, 2009, 13(6): 391-399.

[6] Wernke M, Skvortsov P, Dürr F, et al. A classification of location privacy attacks and approaches[J]. Personal and ubiquitous computing, 2014, 18(1): 163-175.

[7] Wang L, MENG X F. Location privacy preservation in big data era: a survey[J]. Journal of Software, 2014, 25(4): 693-712.

[8] X. Huang, X. Du, "Achieving big data privacy via hybrid cloud," in Proc. of 2014 IEEE INFOCOM Workshops, Pages: 512 - 517

[9] Damiani M L, Cuijpers C. Privacy challenges in third-party location services[C]//Mobile Data Management (MDM), 2013 IEEE 14th International Conference on. IEEE, 2013, 2: 63-66.

[10] Tippenhauer N O, Rasmussen K B, Popper C, et al. Attacks on public WLAN-based positioning systems[C]//Proceedings of the 7th international conference on Mobile systems, applications, and services. ACM, 2009: 29-40.

[11] Li H, Sun L, Zhu H, et al. Achieving privacy preservation in WiFi fingerprint-based localization[C]//INFOCOM, 2014 Proceedings IEEE. IEEE, 2014: 2337-2345.

[12] X. Du, M. Guizani, Y. Xiao and H. H. Chen, "A Routing-Driven Elliptic Curve Cryptography based Key Management Scheme for Heterogeneous Sensor Networks," IEEE Transactions on Wireless Communications, Vol. 8, No. 3, pp. 1223 - 1229, March 2009.

[13] Y. Xiao, V. Rayi, B. Sun, X. Du, F. Hu, and M. Galloway, "A Survey of Key Management Schemes in Wireless Sensor Networks," Journal of Computer Communications, Vol. 30, Issue 11-12, pp. 2314-2341, Sept. 2007.

[14] X. Du, Y. Xiao, M. Guizani, and H. H. Chen, "An Effective Key Management Scheme for Heterogeneous Sensor Networks," Ad Hoc Networks, Elsevier, Vol. 5, Issue 1, pp 24–34, Jan. 2007.

[15] L. Wu, X. Du, and X. Fu, "Security Threats to Mobile Multimedia Applications: Camera-based Attacks on Mobile Phones," IEEE Communications Magazine, Vol. 52, Issue 3, March 2014.

[16] L. Wu, and X. Du, "MobiFish: A Lightweight Anti-Phishing Scheme for Mobile Phones," in Proc. of the 23rd International Conference on Computer Communications and Networks (ICCCN), Shanghai, China, August 2014.

[17] Bomze I M, Budinich M, Pardalos P M, et al. The maximum clique problem[M]//Handbook of combinatorial optimization. Springer, Boston, MA, 1999: 1-74.

[18] Bahl P, Padmanabhan V N. RADAR: An in-building RF-based user location and tracking system[C]//INFOCOM 2000. Nineteenth Annual Joint Conference of the IEEE Computer and Communications Societies. Proceedings. IEEE. Ieee, 2000, 2: 775-784.

[19] Youssef M A, Agrawala A, Shankar A U. WLAN location determination via clustering and probability distributions[C]//Pervasive Computing and Communications, 2003.(PerCom 2003). Proceedings of the First IEEE International Conference on. IEEE, 2003: 143-150.